## Data article

**Title:** Data sets and trained neural networks for Cu migration barriers

**Authors:** Jyri Kimari[1], Ville Jansson[1], Simon Vigonski[1,2], Ekaterina Baibuz[1], Roberto Domingos[3], Vahur Zadin[2], Flyura Djurabekova[1]

**Affiliations:**

[1] Helsinki Institute of Physics and Department of Physics, P.O. Box 43 (Pietari Kalmin katu 2), FI-00014 University of Helsinki, Finland

[2] Institute of Technology, University of Tartu, Nooruse 1, 50411 Tartu, Estonia

[3] Instituto Politécnico de Nova Friburgo – Universidade do Estado do Rio de Janeiro, Rua Sao Francisco Xavier, 524, 20550–900, Rio de Janeiro, RJ, Brazil

**Contact email:** jyri.kimari@helsinki.fi

### Abstract

Kinetic Monte Carlo (KMC) is an efficient method for studying diffusion. A limiting factor to the accuracy of KMC is the number of different migration events allowed in the simulation. Each event requires its own migration energy barrier. The calculation of these barriers may be unfeasibly expensive. In this article we present a data set of migration barriers on for nearest-neighbour jumps on the Cu surfaces, calculated with the nudged elastic band (NEB) method and the tethering force approach. We used the data to train artificial neural networks (ANN) in order to predict the migration barriers for arbitrary nearest-neighbour Cu jumps. The trained ANNs are also included in the article. The data is hosted by the CSC IDA storage service.

**Specifications Table**

| | |
|---|---|
| Subject area | *Physics* |
| More specific subject area | *Kinetic Monte Carlo simulations on the surface, assisted by machine learning* |
| Type of data | *Table* |
| How data was acquired | *Nudged elastic band method with semi-empirical potentials, Cascade2 and iRPROP algorithm for obtaining the neural network weights* |
| Data format | *Raw* |
| Data accessibility | *CSC IDA storage, permanent link http://urn.fi/urn:nbn:fi:att:a4120c3b-9535-405d-b768-4a972bce0b1b* |

**Value of the data**
- The migration barrier tables can be used as a starting point for a Cu surface parameterization for atomistic rigid lattice kinetic Monte Carlo simulations or used as is with some limitations. The data set only contains barriers for the most stable migration events – approximately 17 % of the entire configuration space. In simulations without sharp surface features this barrier set may be sufficient.
- The barriers can serve as a data set in the supervised training of any machine learning method; the trained neural networks can be taken as a comparison point for assessing accuracy.
- The trained neural networks can be used as Cu surface parameterization for kinetic Monte Carlo simulations.

**Data**

Two kinds of data are presented in this article. Firstly, there are the tables of migration energy barriers for various nearest-neighbour jumps on Cu surface. Secondly, there are the presentations of the artificial neural networks trained on the barrier data described above.

The barrier data is in three files: Cu_20160906_100.26d, Cu_20160906_110.26d and Cu_20160906_111.26d. The basis for separating the data in these files is the specific surface on which each migration barrier was calculated – the {100}, the {110} or the {111} surface. These sets were treated as separate data sets when training the neural networks in this article, but in principle the data can be combined if desired for some purpose. The encoding of the local atomic environment (LAE) is equivalent in each subset (see Fig. 1). The format is such that on each line there is the 26-dimensional (26D) encoding as a string of 0s and 1s – 0 denotes a vacancy, 1 denotes a Cu atom, with an ordering corresponding to the indices shown in Fig. 1 – followed by the associated migration energy barrier in eV.

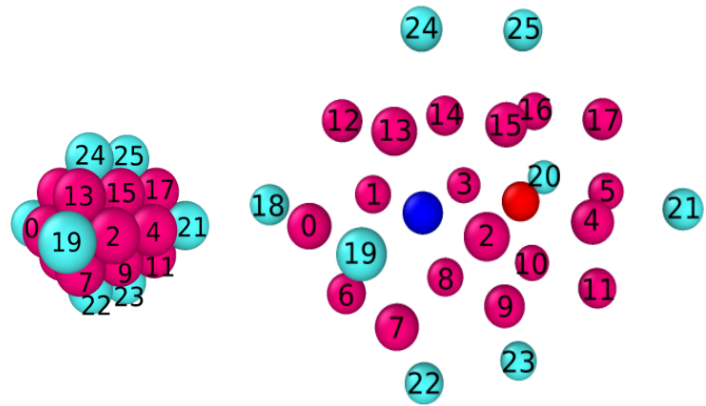

*Figure 1: Encoding of the 26D LAE of a nearest-neighbour jump. The left-hand side displays the lattice sites as touching spheres, the right-hand side shows the sites spread out for clarity. The atom, coloured blue in the right-hand side image, jumps to the lattice site, coloured red. The purple sites are the first-nearest neighbours and the light blue sites are the second-nearest neighbours of the initial or the final position of the jumping atom. The indexing of the 26 neighbour sites from 0 to 25 is shown.*

The neural networks are saved in the format defined by the Fast Artificial Neural Networks (FANN) library [1]. They can be loaded with the fann_create_from_file() function. The loading functionality has been implemented in the Kimocs KMC program [2]; the network files are ready to be used with Kimocs. In total, sixteen different artificial neural networks are presented. Fifteen are regressors that have a 26-dimensional input (the encoded LAE) and a 1-dimensional output (the migration energy in eV). These networks are saved in files Cu_20180209_<100, 110 or 111>_<1, 2, 3, 4 or 5>.net. Five networks correspond to each of the surfaces ({100}, {110}, and {111}). The sixteenth network is a classifier that has 26D input and 3D output – each of the three outputs correspond to the (unnormalized) likelihood of the input LAE to belong in each of the surfaces This network is saved in Cu_20171023_classifier.net. See Fig. 2 for a flowchart describing how to use these sixteen neural networks in a KMC simulation. This is the procedure followed in the KMC simulations of ref. [3]. Obtaining a migration energy barrier from the combination of artificial neural networks during KMC. The transformation of the LAE refers to the procedure where the representative case from the symmetrically equivalent family is chosen. See text and table 1 for details.

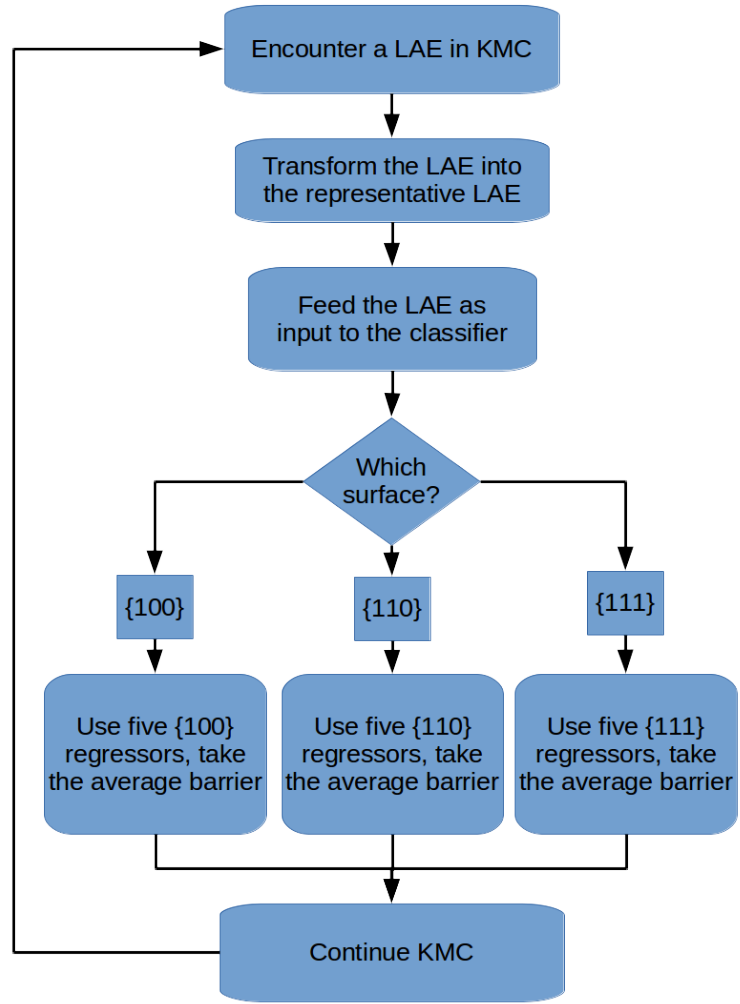

*Figure 2: Obtaining a migration energy barrier from the combination of artificial neural networks during KMC. The transformation of the LAE refers to the procedure where the representative case from the symmetrically equivalent family is chosen. See text and table 1 for details.*

**Computational methods**

The migration barriers have been calculated with the nudged elastic band (NEB) method [4,5] implemented in the LAMMPS molecular dynamics program [6]. The potential energy function was described by the molecular dynamics/Monte Carlo corrected effective medium (MD/MC-CEM) potential by Stave et al. [7]. In addition to the interatomic potential, a tethering force was applied to each atom during NEB; see [8] for details. The force constant of the tethering spring was 2.0 eV/Å$^2$. The simulation box size was approximately 45 Å by 45 Å by 25 Å, but the exact dimensions varied according to the surface orientation used in each calculation. The box beyond the LAE was filled with Cu, so that the LAE was embedded on an open surface. The two bottom atomic layers of the box were fixed. Eleven replicas were used in the NEB calculations. For further details, see ref. [3].

The regressor ANNs are cascade multilayer perceptron (MLP) networks with 30-70 hidden nodes. The activation functions of the hidden nodes are sigmoids, and the activation function at the output node is linear. Training was done for until desired accuracy of 0.08 meV (in the case of the {100} and the {110} data sets) or network size of 70 nodes (in the case of the {111} data set} was reached. The classifier ANN is a fully connected MLP network with 30 hidden nodes in one hidden layer, trained with the iRPROP algorithm.

Prior to training, the barrier data sets were modified by removing the symmetrically redundant LAEs and barriers. The LAE has two planes of symmetry parallel to the jump direction (see Fig. 1), so up to four different LAE encodings correspond to the same physically equivalent environment. There is a third plane of symmetry that is perpendicular to the jump direction and will thus change the barrier. See table 1 for the exact specification of the symmetry operations. The symmetry reduction for the regressors was done so that from each group of up to four equivalent LAEs the encoding with the *smallest value* was selected as the representative to remain in the data set. The value is defined as the numeric value of the 26-dimensional vector when its elements are concatenated to a single number. For training the classifier, the representative was taken from the entire family of up to eight LAEs; including the reverse processes. This way, the forward and the reverse process barriers will always be taken from the same regressor, despite the quality of the classifier.

*Table 1: Symmetry operations associated with each plane of symmetry. Horizontal and vertical refer to the directions as shown in Fig. 1. The planes parallel to the jump direction preserve the jump direction, leaving the barrier unchanged; mirroring perpendicular to the jump direction changes the barrier.*

| Symmetry plane | Parallel to the jump direction, "horizontal" | Parallel to the jump direction, "vertical" | Perpendicular to the jump direction |
|---|---|---|---|
| Element pairs to be swapped | 6 – 12, 7 – 13, 8 – 14, 9 – 15, 10 – 16, 11 – 17, 22 – 24, 23 – 25 | 1 – 2, 3 – 4, 6 – 7, 8 – 9, 10 – 11, 12 – 13, 14 – 15, 16 – 17, 18 – 19, 20 – 21 | 0 – 5, 1 – 3, 2 – 4, 6 – 10, 7 – 11, 12 – 16, 13 – 17, 18 – 20, 19 – 21, 22 – 23, 24 – 25 |

The symmetries have to be taken into account when using the network. This is marked in the flowchart of Fig. 2 as "Transform the LAE into the representative LAE". This means that whenever a LAE is encountered in KMC, the eight mirror images of that LAE are first generated, and then the one with the smallest value is selected and given as input to the classifier instead of the original LAE. After classification, the smallest image *with the same jump direction* (not considering the reverse processes in this stage) is given as input for the appropriate regressors.

**Acknowledgements**
J. Kimari was supported by a CERN K-contract and Academy of Finland (Grant No. 313867). V. Jansson was supported by Academy of Finland (Grant No. 285382). F. Djurabekova acknowledges gratefully the financial support of Academy of Finland (Grant No. 269696) and MEPhI Academic Excellence Project (Contract No. 02.a03.21.0005). E. Baibuz was supported by a CERN K-contract and the doctoral program MATRENA of the University of Helsinki. V. Zadin and S. Vigonski were supported by Estonian Research Council grant PUT 1372. Computing resources were provided by the Finnish IT Center for Science (CSC) and the Finnish Grid and Cloud Infrastructure (persistent identifier urn:nbn:fi:research-infras-2016072533).